\documentclass[letterpaper,11pt]{article}
\pdfoutput=1 

\usepackage{jcappub} 

\usepackage[T1]{fontenc}
\usepackage{booktabs} 
\usepackage{subcaption} 
\usepackage{caption}
\usepackage{orcidlink}

\usepackage{siunitx} 
\usepackage{mathtools} 
\usepackage{textcomp} 

\usepackage{parskip}

\hypersetup{
    colorlinks=true,
    linkcolor=blue,
    filecolor=magenta,      
    urlcolor=blue,
    citecolor=teal
}

\usepackage{cleveref} 

\usepackage{xcolor}
\pagecolor{white} 

\title{Bayesian analysis of $\alpha$-Starobinsky model with Planck, ACT and DESI data}

\author[a,c]{Karim Carrion\orcidlink{0000-0002-1798-7978},}
\author[b]{Francisco X. Linares Cede\~no\orcidlink{0000-0002-9733-2209},}
\author[a]{Gabriel Germ\'an\orcidlink{0000-0001-8597-9796},}
\author[a]{Juan Carlos Hidalgo\orcidlink{0000-0001-9715-1232}\,}

\affiliation[a]{Instituto de Ciencias F\'{\i}sicas, Universidad Nacional
Aut\'onoma de M\'exico,\\ Av. Universidad s/n, Cuernavaca, Morelos, 62210, M\'exico.}
\affiliation[b]{Facultad de Ciencias Físico-Matemáticas, Universidad Michoacana de San Nicolás de Hidalgo,
Edificio Alfa, Ciudad Universitaria, 58040 Morelia, Michoacán, México,}
\affiliation[c]{Instituto de Física, Universidad Nacional Autónoma de México, \\ Circuito de la Investigación Científica, Ciudad Universitaria, Cd. de México C. P. 04510, México. \\}

\emailAdd{kcarrion@fisica.unam.mx}
\emailAdd{francisco.linares@umich.mx}
\emailAdd{gageve@gmail.com}
\emailAdd{hidalgo@icf.unam.mx}

\abstract{We present a joint Bayesian analysis to impose constraints on the generalized $\alpha$-Starobinsky inflationary model using the high-precision cosmological datasets: Planck, CMB lensing from ACT DR6, and Baryon Acoustic Oscillations (BAO) from DESI DR2. For the parameter inference, we introduce an alternative sampling approach. Rather than imposing priors on the cosmological parameters of the inflationary potential $(V_0, \, \alpha, \, N_*)$, we place priors directly on the primordial physical observables $(A_s,\, n_s,\, r)$ through analytical slow-roll consistency relations. Our pipeline internally maps these sampled observables to the corresponding $\alpha$-Starobinsky parameters. These values are then passed to a modified version of \texttt{CLASS}, which solves the exact inflationary dynamics numerically. This pipeline ensures that the final reported posteriors for the observables are computed exactly, completely free from the slow-roll approximation. Applying this methodology, we explore the viability of the $\alpha$-Starobinsky model. We show that, when the full combined dataset is considered, the pure Starobinsky model (i.e., the canonical limit $\alpha = 1$) shows signs of breaking down, since it requires a large number of e-folds of inflation after horizon crossing ($N_* > 60$) due to the shift in the scalar spectral index, $n_s$. In contrast, allowing the deformation parameter $\alpha$ as a free parameter yields a clear $1\sigma$ preference for $\log_{10} \alpha > 0$, present across all datasets. Notably, we also show that the addition of ACT DR6 lensing data introduces no significant impact on these primordial constraints, confirming that our robust posteriors are primarily driven by Planck and DESI measurements. 
}

\begin{document}
\maketitle
\flushbottom

\section{Introduction}
\label{sec:intro}

Cosmic inflation currently stands as the cornerstone paradigm for the early Universe, elegantly resolving the horizon, flatness, and monopole problems of standard Big Bang cosmology, while providing a quantum mechanical origin for the primordial density perturbations that seeded large-scale structure (for reviews on inflation, see e.g., \cite{1984RPPh...47..925L,1999PhR...314....1L,2009arXiv0907.5424B,2013arXiv1303.3787M,2017IJMPD..2640002B,2022arXiv220308128A,2025GReGr..57..135K}). The simplest models of inflation, driven by a single canonical scalar field slowly rolling down a nearly flat potential, predict a nearly scale-invariant, adiabatic, and highly Gaussian spectrum of primordial perturbations. Far from remaining purely theoretical, these core predictions have been rigorously validated by successive generations of Cosmic Microwave Background (CMB) experiments \cite{2013ApJS..208...20B,2020A&A...641A...1P,2024PhRvD.109f3530C,2023PhRvD.108b3510B}. The culmination of this effort, the Planck satellite's legacy data release \cite{2020A&A...641A...6P, Planck2018_inflation}, now provides the most precise benchmark measurements of the CMB anisotropies that ruthlessly tests the viability of competing early-universe scenarios, effectively ruling out several previously viable models. 

Among the plethora of inflationary models developed over the years \cite{2024PDU....4601653M}, the Starobinsky model \cite{1980PhLB...91...99S} has emerged as a particularly compelling candidate, predating even the formalization of the cosmic inflation paradigm. Formulated as a geometric modification of General Relativity, featuring a quadratic Ricci scalar term ($R^2$), it can be conformally mapped to the Einstein frame as a canonical scalar field rolling along a plateau-like potential. Planck data has severely constrained simple monomial potentials, showing a strong preference for exactly this type of asymptotically flat plateau, which naturally predicts a small tensor-to-scalar ratio $r$ and a scalar spectral index $n_s$ in excellent agreement with observations \cite{Planck2018_inflation}. To precisely embed such phenomenological successes into theories such as supergravity, a broader theoretical framework known as $\alpha$-attractors has been developed \cite{2013JCAP...07..002K, 2013PhRvD..88h5038F}. Directly emerging from this class of model is the $\alpha$-Starobinsky model (or so-called E-model). Extensively studied in recent years \cite{2013JCAP...10..009E,2013JHEP...11..198K,2019JCAP...09..040E,2025JCAP...01..116B}, this scenario introduces a tuning parameter $\alpha$ that is inversely proportional to the scalar curvature of the internal K\"ahler manifold \cite{2000CQGra..17.4269K}. This parameter generalizes the standard Starobinsky prediction ($\alpha=1$), modulating the width of the inflationary plateau and offering a flexible, yet theoretically rigorous, parameter space.

While Planck provides stringent constraints on this parameter space, the advent of next-generation cosmological surveys offers a crucial opportunity to test these models under new observational stress. High-resolution CMB measurements from the Atacama Cosmology Telescope (ACT) provide refined insights into small-scale temperature and polarization anisotropies, as well as late-time structure growth via CMB lensing \cite{2024ApJ...962..113M,2024ApJ...962..112Q}. Simultaneously, the recent Data Release 2 (DR2) from the Dark Energy Spectroscopic Instrument (DESI) has mapped the cosmic expansion history, at lower redshifts, with unprecedented precision through Baryon Acoustic Oscillations (BAO) \cite{2025PhRvD.112h3515A}. The inclusion of these independent datasets, particularly the DESI BAO measurements, visibly shifts the preferred values of standard cosmological parameters like $n_s$ \cite{2025JCAP...11..063C,2025JCAP...11..062L}. For highly rigid models like the pure Starobinsky case ($\alpha=1$), accommodating these shifts requires an undesirably large number of large number of $e$-folds during inflation, $N_*$. This substantial inconsistency directly motivates the exploration of the $\alpha$-deformed parameter space, where the additional degree of freedom allows the model to naturally absorb the statistical shift of modern datasets. 

In this work, we perform a comprehensive Bayesian analysis of the $\alpha$-Starobinsky model using datasets from Planck, ACT, and DESI. A key methodological feature of our approach is the parameterization of the primordial priors. Rather than sampling the inherently degenerate theoretical parameters of the potential ($V_0, \alpha, N_*$), we sample directly from the observationally constrained physical observables ($A_s^\alpha, n_s^\alpha, r^\alpha$) using analytical consistency relations valid in a slow-roll regime (see e.g. \cite{2015PhRvD..91f4016M,2023JCAP...12..015G}). These sampled points act as dynamical seeds to reconstruct the exact potential parameters, which is then fed into an amended version of the Boltzmann solver \texttt{CLASS} \cite{2011arXiv1104.2932L,2011JCAP...07..034B}, then evolves the full background and perturbed field equations natively.
Consequently, the final constraints are entirely free from slow-roll approximations, ensuring numerically rigorous posteriors from the Boltzmann code.

The remainder of this paper is structured as follows. In \autoref{sec:alpha-staro}, we review the theoretical formulation of the $\alpha$-Starobinsky model and derive the consistency relations that enable our parameterization. Then, \autoref{sec:methodology} details our numerical methodology, including the specific modifications made to \texttt{CLASS} and the effects on observables. Furthermore, it outlines the cosmological likelihoods employed in our Markov Chain Monte Carlo (MCMC) analysis. The core observational constraints are presented in \autoref{sec:constraints}, where we contrast the rigid Starobinsky ($\alpha=1$) against the generalized $\alpha$-deformed case. Finally, we provide our concluding remarks in \autoref{sec:conclusions}. Supplementary visualizations that have been omitted from the main text are provided in \autoref{app:A} and \autoref{app:B}. 

\section{The \texorpdfstring{$\alpha$}{alpha}-Starobinsky model}
\label{sec:alpha-staro}

The Starobinsky model \cite{1980PhLB...91...99S} represents one of the earliest successful formulations of an inflationary phase. Its unique feature is the modification of the Einstein-Hilbert action through the addition of a quadratic scalar curvature term, expressed as

\begin{equation} 
S = \frac{M_{\rm Pl}^{2}}{2} \int d^{4}x \sqrt{-g} \left( R + \frac{1}{6M^2} R^{2} \right) \, ,
\label{eq:Staro_action} 
\end{equation} 

where $R$ denotes the Ricci scalar, $g$ is the determinant of the metric tensor $g_{\mu\nu}$, $M$ represents a characteristic mass scale, and $M_{\rm Pl} \simeq 2.44 \times 10^{18} \, \mathrm{GeV}$ is the reduced Planck mass. Although formulated in terms of higher-order gravity, the action can be recast into a more intuitive scalar-field representation via a conformal transformation of the metric. This procedure maps the theory into the Einstein frame, yielding the action

\begin{equation} 
S = \int d^{4}x \sqrt{-g} \left( \frac{M_{\rm Pl}^{2}}{2} R - \frac{1}{2} \partial_{\mu}\phi \partial^{\mu}\phi - V(\phi) \right) \, . 
\label{eq:full_action}
\end{equation} 

Under this transformation, the degree of freedom associated with the $R^2$ term manifests as a canonical scalar field $\phi$ evolving within a plateau-like potential

\begin{equation} 
V(\phi) = V_{0} \left[ 1 - \exp \left( -\sqrt{\frac{2}{3}} \frac{\phi}{M_{\rm Pl}} \right) \right]^{2} \, , 
\label{eq:Staro_pot} 
\end{equation} 

where the vacuum energy scale is defined by $V_{0} = \frac{3}{4} M_{\rm Pl}^{2} M^2$ in units of $M_{\rm Pl}^4$. Once the model is framed within the Einstein representation and the scalar potential $V(\phi)$ is explicitly defined, we can apply the standard slow-roll approximation for single-field inflation. This approach allows us to derive analytical expressions for the primary cosmological observables, specifically the tensor-to-scalar ratio $r$ and the scalar amplitude $A_{s}$. Furthermore, we can establish the consistency relations for the scalar spectral index $n_{s}$ and the scalar running $n_{sk}$. 

Nonetheless, in this work, we focus on the $\alpha$-Starobinsky model, which represents a straightforward yet non-trivial generalization of the original potential in Eq.~\eqref{eq:Staro_pot} through the introduction of a deformation parameter $\alpha$. Motivated by specific frameworks in supergravity \cite{2013JCAP...10..009E,2013JHEP...11..198K,2019JCAP...09..040E}, the generalized potential is defined as
\begin{equation}
V(\phi) = V_{0} \left[ 1 - \exp \left( -\sqrt{\frac{2}{3\alpha}} \frac{\phi}{M_{\rm Pl}} \right) \right]^{2} \, .
\label{eq:alpha_starobinsky}
\end{equation}

\begin{figure}[htp]
\centering
  \includegraphics[width=\linewidth]{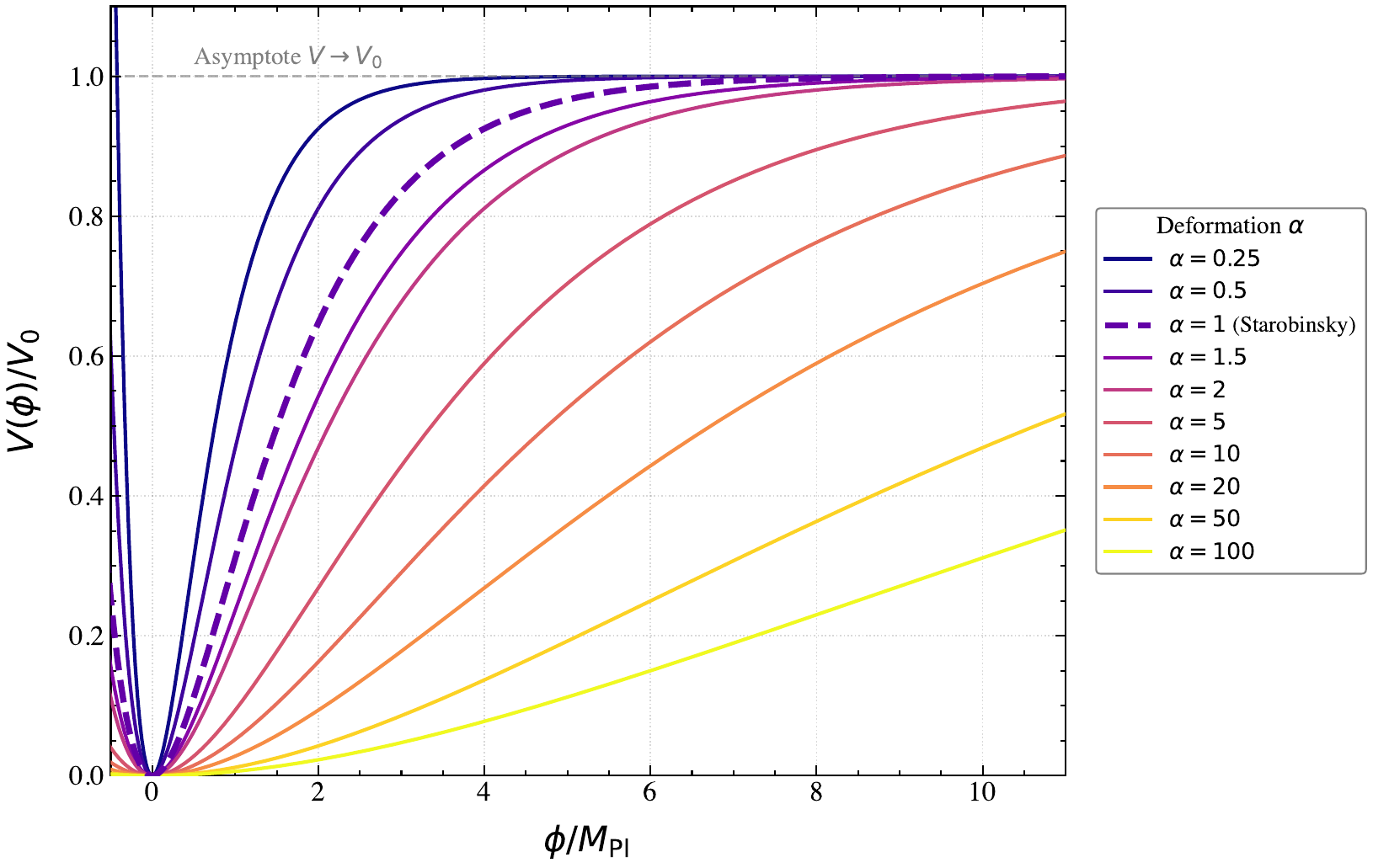}
  \caption{Dimensionless inflationary potential, $V(\phi)/V_0$, for the $\alpha$-Starobinsky model in Eq.~\eqref{eq:alpha_starobinsky}. Colored curves indicate different values of the deformation parameter $\alpha$, interpolating between a sharper rise at small field values ($\alpha=0.25$) and an extended, flatter plateau ($\alpha=100$). The dashed curve marks the Starobinsky limit $\alpha=1$. Increasing $\alpha$ flattens the potential and delays its approach to the asymptotic value $V \rightarrow V_0$, effectively widening the inflationary plateau.}
  \label{potalfa}
\end{figure}

This class of potentials is particularly compelling as it preserves the observational viability of the original model while offering a broader parameter space for theoretical exploration. The influence of the parameter $\alpha$ on the inflationary landscape is illustrated in \autoref{potalfa}, where it can be seen that $\alpha$ modulates the curvature of the potential and the width of the inflationary plateau. 

\subsection{A supergravity framework}
\label{subsec:sugra}

Within the framework of supergravity, the inflationary dynamics are fundamentally determined by the K{\"a}hler potential $K$ and the superpotential $W$. For minimal supergravity configurations, the K{\"a}hler potential is typically defined as $K = \phi^{i}\phi_{i}^{*}$, where $\phi_{i}$ denotes the chiral scalar fields of the theory. Working in reduced Planck units ($M_{\rm Pl}=1$), the resulting effective potential of the scalar field can be derived as

\begin{equation}
V = e^{K} \left[ K^{i\bar{j}} D_i W \overline{D_j W} - 3|W|^{2} \right] \, ,
\label{eq:Superpotential}
\end{equation}

where $D_i W = \partial_{\phi_i} W + (\partial_{\phi_i} K) W$ is the K{\"a}hler derivative. A well-known issue in minimal supergravity is that the $-3|W|^2$ term prevents the potential from being positive definite, often resulting in Anti-de Sitter (AdS) minima that are unsuitable for inflation.

To circumvent this, one appeals to no-scale supergravity \cite{1983PhLB..133...61C,1987PhR...145....1L}, where the specific logarithmic structure of the K{\"a}hler potential leads to a cancellation of the negative term in Eq.~\eqref{eq:Superpotential}, naturally yielding a positive semi-definite flat potential. The no-scale K{\"a}hler potential is defined as

\begin{equation}
K = -3 \log \left( T + T^* - \frac{|\phi_{i}|^{2}}{3} \right) \, ,
\label{Eq:Kahler_potential}
\end{equation}

where $T$ represents the volume modulus. By stabilizing $T$ and employing the Wess-Zumino superpotential \cite{2013PhRvL.111k1301E}

\begin{equation}
W(\phi) = \mu \left( \frac{1}{2}\phi^2 - \frac{1}{3\sqrt{3}}\phi^3 \right) \, ,
\label{Eq:WZ_superpotential}
\end{equation}

the Starobinsky-like plateau is recovered. To generalize this framework to the $\alpha$-Starobinsky model, the K{\"a}hler potential is extended to

\begin{equation}
K = -3\alpha \log \left( T + T^* - \frac{|\phi_{i}|^{2}}{3} \right) \, .
\label{eq:no-scale_alpha}
\end{equation}

While this generalization requires a more sophisticated superpotential to ensure stability \cite{2019JCAP...09..040E}, it provides a clear geometric interpretation: the parameter $\alpha$ is inversely proportional to the scalar curvature of the underlying internal K{\"a}hler manifold.

\subsection{Consistency relations}
\label{subsec:consistency_relations}

As we mentioned, the connection with the inflationary landscape is established through cosmological observables evaluated under the slow-roll approximation. To first order in the slow-roll parameters, these observables are defined by the following set of relations \cite{1994PhRvD..50.7222L,1999PhR...314....1L}

\begin{align}
n_{t} &= -2\epsilon = -\frac{r}{8} \, , \label{eq:Int} \\
n_{s} &= 1 + 2\eta - 6\epsilon \,  , \label{eq:Ins} \\
n_{sk} &\equiv \frac{d n_s}{d \log k} = 16\epsilon \eta - 24\epsilon^{2} - 2\xi_2 \,  , \label{eq:Insk} \\
A_s(k) &= \frac{1}{24\pi^{2}} \frac{V}{\epsilon M_{\rm Pl}^{4}}\, . \label{eq:IAs} 
\end{align}

Here, the amplitude of density perturbations at wave number $k$ is denoted by $A_s(k)$ and $n_{sk}$ is the running of the scalar index, usually denoted by $\alpha$. All these quantities are evaluated at horizon crossing for the given wave number $k$. The slow-roll parameters mentioned above, which are defined in terms of the potential $V(\phi)$ and its derivatives, are defined as

\begin{equation}
\epsilon \equiv \frac{M_{\rm Pl}^{2}}{2} \left( \frac{V'}{V} \right)^{2}, \quad \eta \equiv M_{\rm Pl}^{2} \frac{V''}{V}, \quad \xi_2 \equiv M_{\rm Pl}^{4} \frac{V' V'''}{V^{2}} \, .
\label{Eq:slowroll_params}
\end{equation}

In which, the primes denote successive derivatives with respect to the inflaton field $\phi$. Considering the $\alpha$-Starobinsky potential in Eq.~\eqref{eq:alpha_starobinsky} and then evaluating the condition $16\epsilon = r$ at horizon crossing, we can determine the field value $\phi_k$ in terms of the tensor-to-scalar ratio and the deformation parameter

\begin{equation}
\phi_{k} = \sqrt{\frac{3\alpha}{2}} \log \left( 1 + \frac{8}{\sqrt{3\alpha r}} \right) \, .
\label{Eq:phi_k}
\end{equation}

Substituting the potential and its derivatives into the expression for the scalar spectral index, we establish a fundamental consistency relation between $r$, $n_s$, and $\alpha$

\begin{equation}
r = \frac{4 \left( 2 + 3 \alpha \delta_{n_s} - 2 \sqrt{1 + 3 \alpha \delta_{n_s}} \right)}{3 \alpha} \, ,
\label{r1}
\end{equation}

where we define $\delta_{n_s} \equiv 1 - n_s$. Inverting this relation allows us to express the model parameter $\alpha$ solely in terms of the primary cosmological observables

\begin{equation}
\alpha = \frac{16r}{3(4\delta_{n_s} - r)^2} \, .
\label{eq:consistency_1}
\end{equation}

Furthermore, the overall normalization of the potential, $V_0$, is determined by solving the scalar perturbation amplitude Eq.~\eqref{eq:IAs} at horizon crossing. This yields

\begin{equation}
V_0 = \frac{3 A_s \pi^2 r (8\delta_{n_s} - r)^2}{8(4\delta_{n_s} - r)^2} M_{\rm Pl}^4 \, ,
\label{eq:consistency_2}
\end{equation}

where $A_s$ is the scalar amplitude evaluated at the pivot scale $k_* = 0.05\,\mathrm{Mpc}^{-1}$ (with a fiducial value of $2.1 \times 10^{-9}$). In a similar manner, we derive the consistency relations for the scalar running

\begin{equation}
n_{sk} = -\frac{1}{64}(8\delta_{n_s} - r)(4\delta_{n_s} + r) \, . \label{eq:nsk1} 
\end{equation}

Finally, the number of $e$-folds of inflationary expansion, $N_k$, can be expressed as a function of the model parameters {{}by evaluating the slow-roll integral $N_k = \frac{1}{M_{\rm Pl}^2} \int_{\phi_{\rm end}}^{\phi_k} \frac{V}{V'}\,d\phi$ for the $\alpha$-Starobinsky potential and rewriting the result in terms of $r$ via $r = 16\epsilon$:}

\begin{equation}
N_{k} = \frac{3 \alpha}{4} \left[ \frac{2(4 - \sqrt{r})}{\sqrt{3\alpha r}} + \log \left( \frac{\sqrt{r}(2 + \sqrt{3\alpha})}{8 + \sqrt{3\alpha r}} \right) \right].
\label{eq:alphaNk}
\end{equation}

By eliminating $\alpha$ via Eq.~\eqref{eq:consistency_1}, we finally obtain $N_k$ (denoted as $N_*$ at the pivot scale) purely as a function of the observables

\begin{equation}
N_{*} = \frac{2(4 - \sqrt{r})}{4\delta_{n_s} - r} + \frac{4r}{(4\delta_{n_s} - r)^2} \log \left[ \frac{\sqrt{r}(4\delta_{n_s} - r + 2\sqrt{r})}{2(8\delta_{n_s} - r)} \right] \, .
\label{eq:consistency_3}
\end{equation}

\section{Methodology and data}
\label{sec:methodology}

A key methodological contribution of this work is the application of the consistency relations {{} given by Eqs.~}\labelcref{eq:consistency_1,eq:consistency_2,eq:consistency_3} to strategically parametrize the sampling process. Rather than exploring the standard $\alpha$-Starobinsky set $\{V_0, \alpha, N_{*}\}$ of parameters, we transform them to sample directly from the well-known observational variables $\{A^\alpha_s, n^\alpha_s, r^\alpha\}$, where the superscript $\alpha$ denotes that these quantities are derived through the aforementioned consistency relations. This procedure, which we denote as \textit{slow-roll parameterization} on priors, is crucial: it allows for the adoption of priors from established pipelines, such as Planck, effectively anchoring the prior volume to observational bounds rather than abstract model scales. By mapping the theoretical requirements of the $\alpha$-Starobinsky framework onto the $\{A^\alpha_s, n^\alpha_s, r^\alpha \}$ space, the Bayesian analysis is naturally directed toward the most physically viable regions of the parameter space. Ultimately, this transformation ensures that the posterior exploration is restricted to the theoretically permitted regimes of the model, providing a more robust and physically motivated inference.

\begin{figure*}[!b]
    \centering
    \begin{subfigure}[b]{0.48\textwidth}
        \centering
        \includegraphics[width=\textwidth]{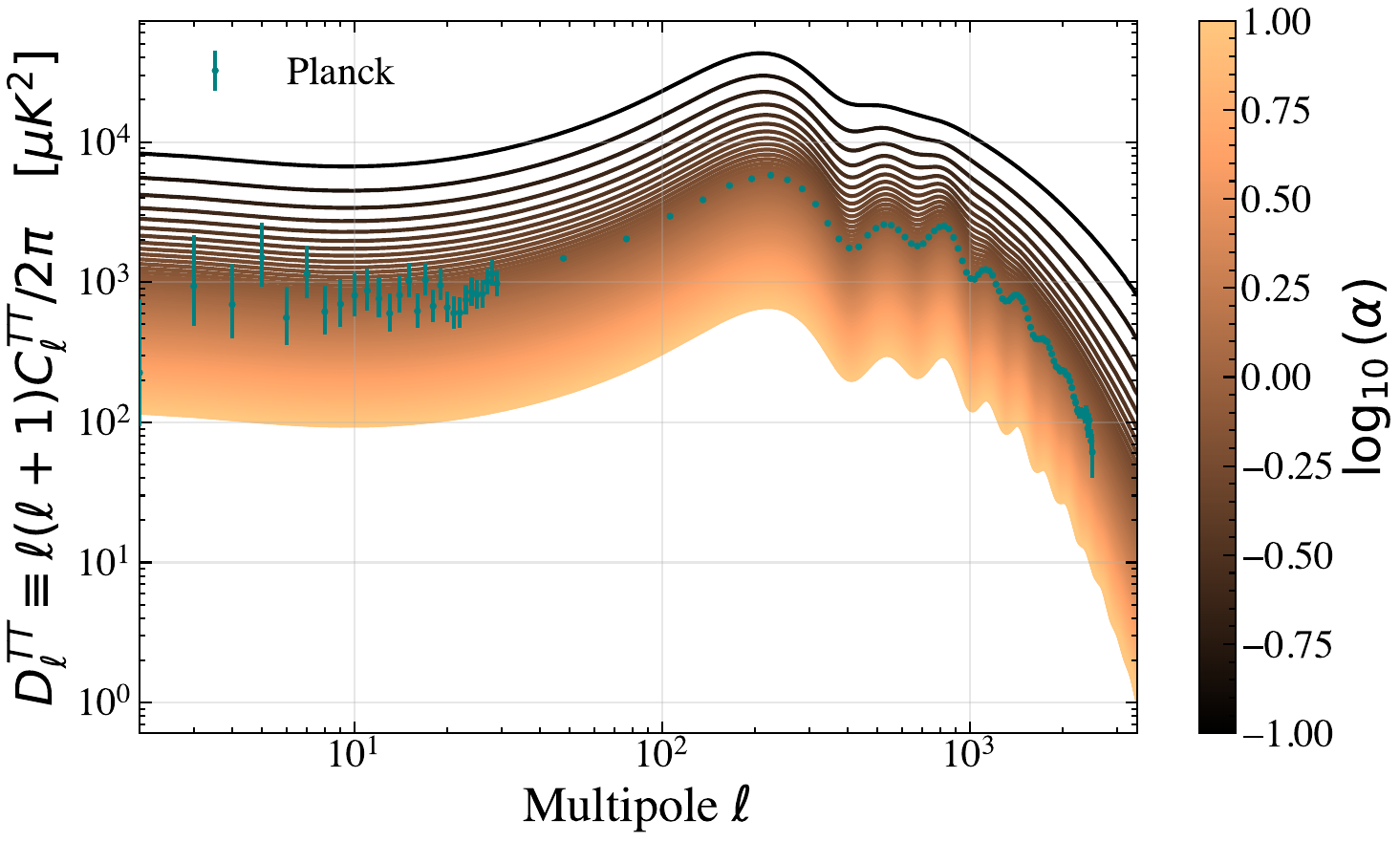}
        \caption{Varying deformation parameter $\alpha$}
        \label{fig:tt_alpha}
    \end{subfigure}
    \hfill
    \begin{subfigure}[b]{0.48\textwidth}
        \centering
        \includegraphics[width=\textwidth]{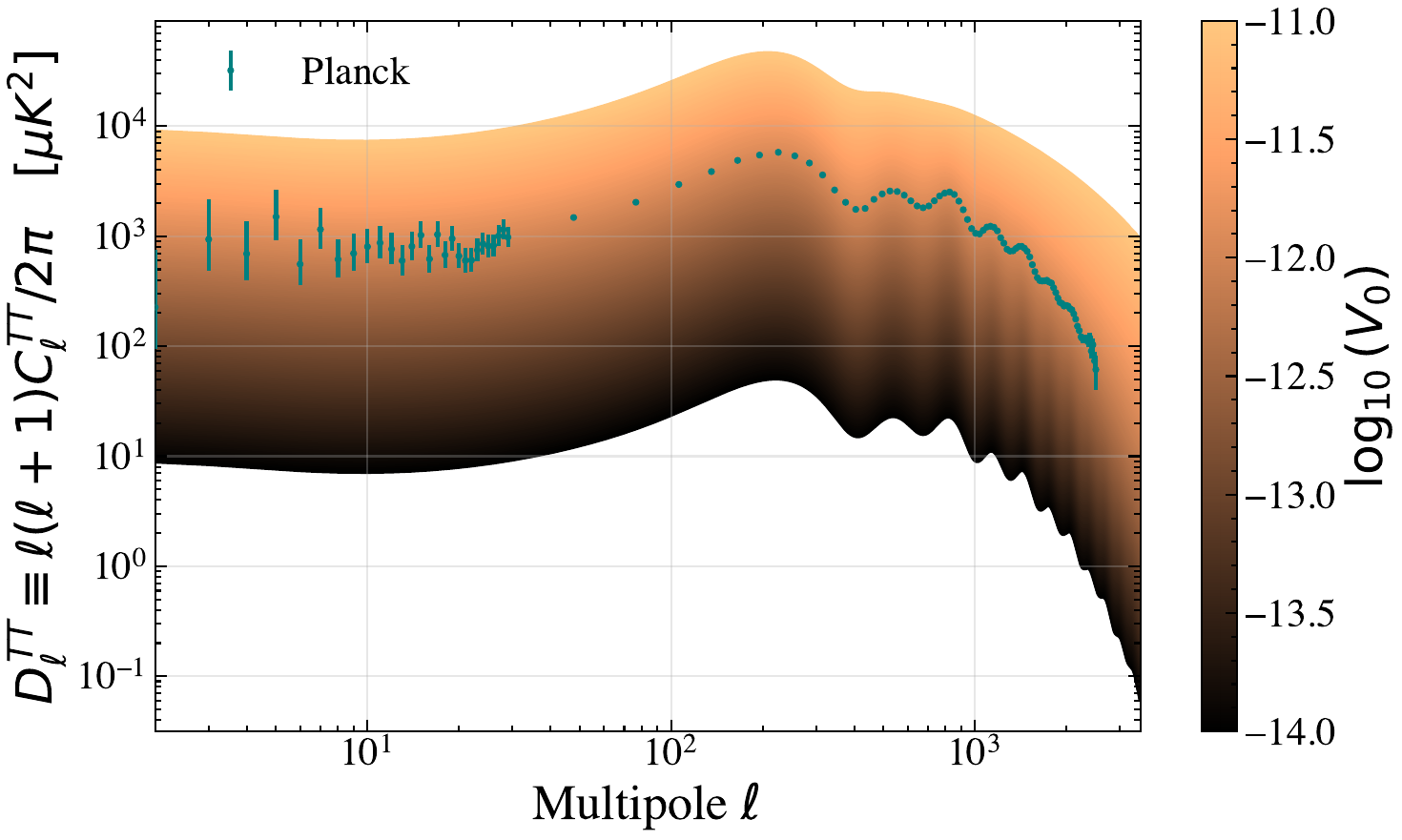}
        \caption{Varying potential scale $V_0$}
        \label{fig:tt_v0}
    \end{subfigure}
    
    \vspace{0.4cm} 
    
    \begin{subfigure}[b]{0.48\textwidth}
        \centering
        \includegraphics[width=\textwidth]{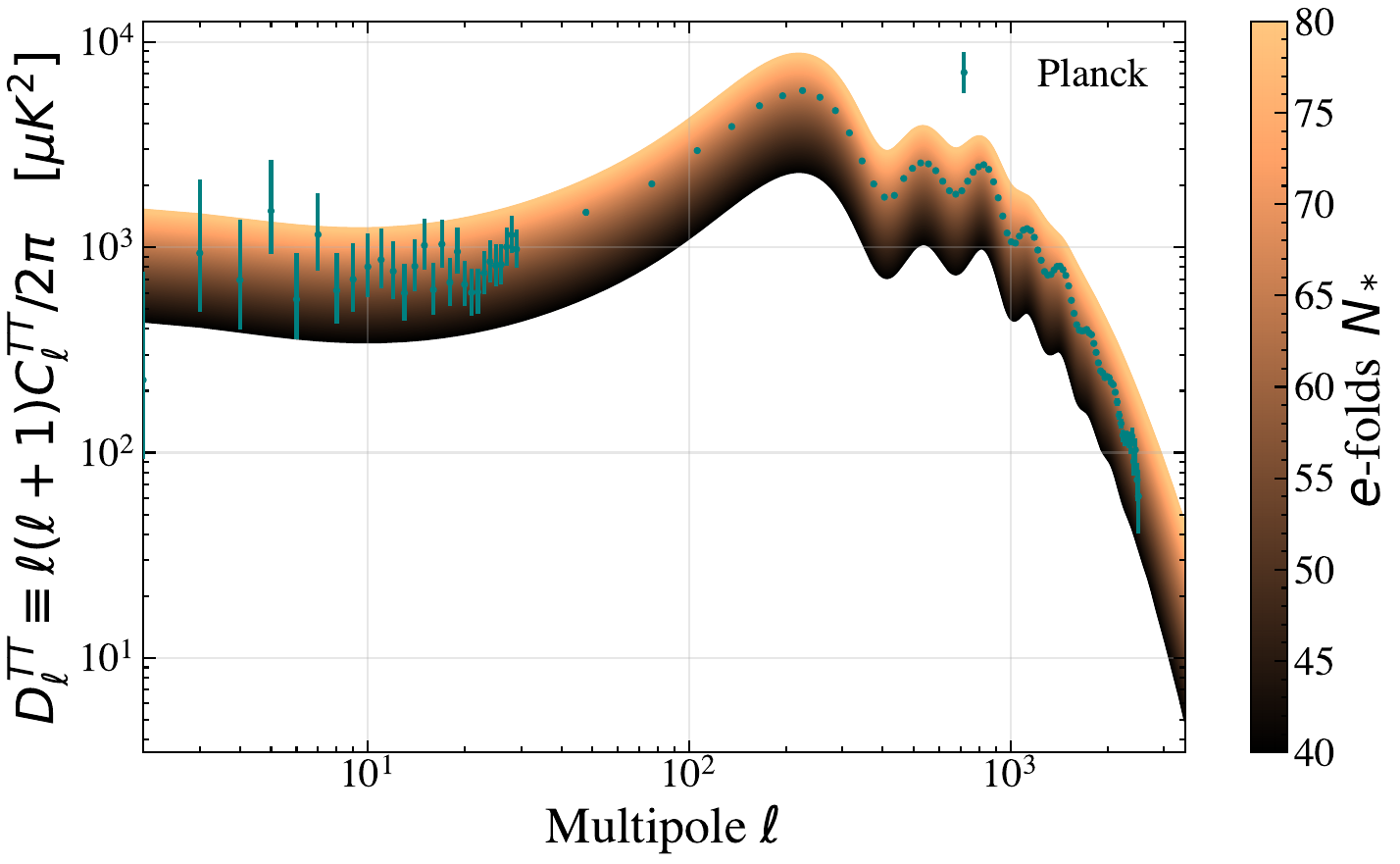}
        \caption{Varying number of $e$-folds $N_*$}
        \label{fig:tt_nstar}
    \end{subfigure}
    
    \caption{CMB temperature anisotropy power spectrum $\mathcal{D}_\ell^{TT} \equiv \ell(\ell+1)C_\ell^{TT}/2\pi$ for the $\alpha$-Starobinsky model. Numerical predictions obtained from our modified Boltzmann solver are compared against Planck 2018 data (teal points). The panels illustrate the impact of systematically varying the model's core parameters: \textbf{(a)} the deformation parameter $\alpha$ (plotted as $\log_{10}\alpha \in [-1, 1]$), \textbf{(b)} the potential scale $V_0$ ($\log_{10}V_0 \in [-14, -11]$), and \textbf{(c)} the number of $e$-folds $N_*$ ($N_* \in [40, 80]$). In each panel, while one parameter is swept across the copper colormap, the other two are held fixed at their fidual values. The prominent vertical dispersion in \textbf{(a)} and \textbf{(b)} illustrates the strong dependence of the overall scalar amplitude ($A_s$) on both $\alpha$ and $V_0$. Despite these large amplitude shifts, the stable horizontal placement of the acoustic peaks confirms that the varying parameters preserve the fundamental acoustic structure without inducing gross scale-dependent deformations.}
    \label{fig:TT_impact}
\end{figure*}

To compute the theoretical predictions for the CMB anisotropies, we modified the primordial module of the Boltzmann solver \texttt{CLASS}\footnote{\href{https://github.com/karimpsi22/class\_alphastaro}{github.com/karimpsi22/class\_alphastaro}} to directly solve the inflationary dynamics for the generalized $\alpha$-Starobinsky potential (E-models) defined in Eq.~\eqref{eq:alpha_starobinsky}. This approach yields the lensed temperature (TT), polarization (EE), and cross-correlation (TE) angular power spectra used within the likelihood analysis. We adopt numerical values expressed in Planck units, defining the reduced Planck mass as $M_{\mathrm{Pl}} = 1/\sqrt{8\pi}$ (consistent with the internal convention of \texttt{CLASS}, where the physical value is $2.44\times 10^{18}\,\mathrm{GeV}$). Consequently, for the primordial power spectrum, we use a pivot scale of $k_* = 1.3128\times 10^{-58}M_{\mathrm{Pl}}$, which corresponds to the standard value of $k_* = 0.05\,\mathrm{Mpc}^{-1}$ used in \texttt{CLASS}. \\ 
In \autoref{fig:TT_impact}, we explore the sensitivity of the CMB temperature anisotropy power spectrum to the main parameters of the $\alpha$-Starobinsky model using our modified \texttt{CLASS}. \autoref{fig:tt_alpha} demonstrates that varying the deformation parameter $\alpha$ induces significant vertical shifts in the spectra. Because the value of potential scale $V_0$ is held fixed in this panel, altering $\alpha$ directly modifies the inflationary energy scale and, consequently, the overall scalar amplitude ($A_s$). A mathematically similar amplitude modulation is observed in \autoref{fig:tt_v0} when sweeping $V_0$ across multiple orders of magnitude while holding $\alpha$ constant, reinforcing that both parameters are tightly coupled to the normalization of the power spectrum. Finally \autoref{fig:tt_nstar} illustrates the effect of varying the duration of inflation via the number of $e$-folds, $N_*$. While $N_*$ also impacts the amplitude, its variation slightly adjusts the scalar spectral index ($n_s$), introducing a subtle tilt to the spectrum. Across all three panels, the consistent positioning of the acoustic peaks confirms that while the parameter choices heavily dictate the amplitude, the fundamental acoustic structure and underlying physics remain robust and well-aligned with the Planck 2018 data.

\begin{table}[htp]
\centering
\renewcommand{\arraystretch}{1.4} 
\setlength{\tabcolsep}{10pt}      

\begin{tabular}{l c c}
\toprule
\textbf{Parameter} & \textbf{Lower Limit} & \textbf{Upper Limit} \\
$\quad \log(10^{10} A^{\alpha}_\mathrm{s})$ & $1.61$ & $3.91$ \\
$\quad n^{\alpha}_\mathrm{s}$ & $0.8$ & $1.2$ \\
$\quad r^{\alpha}$ & $0$ & $3$ \\
$\quad 100 \, \theta_\mathrm{s}$ & $0.5$ & $10$ \\
$\quad \Omega_\mathrm{b} h^2$ & $0.005$ & $0.1$ \\
$\quad \Omega_\mathrm{c} h^2$ & $0.001$ & $0.99$ \\
$\quad \tau_\mathrm{reio}$ & $0.01$ & $0.8$ \\
\bottomrule
\end{tabular}
\caption{Uniform priors applied in the MCMC sampling after parametrizing $\{ V_0,\alpha,N_* \}$ into $\{ \log(10^{10}A_s^\alpha), n_s^\alpha, r^\alpha \} $ using the consistency relations \labelcref{eq:consistency_1,eq:consistency_2,eq:consistency_3}. For the $\alpha=1$ case, the $r^\alpha$ prior is omitted due to the reduced number of inflationary degrees of freedom. All remaining cosmological parameters adopt standard, broad priors consistent with established CMB pipelines {{}(specifically, the Planck 2018 priors as provided by \texttt{Cobaya})}.}
\label{table_priors_adapted}
\end{table}

The Bayesian inference analysis is performed using \texttt{Cobaya}\footnote{\href{https://github.com/CobayaSampler/cobaya}{github.com/CobayaSampler/cobaya}} \cite{2021JCAP...05..057T}. We assign flat priors to each cosmological parameter, as detailed in \autoref{table_priors_adapted}; notably, these prior ranges are adopted from {{}established} observational pipelines {{}(e.g., those from Planck 2018, SPT, and DES)}, ensuring they have been rigorously tested against established values for $A_s$, $n_s$, and $r$. During the sampling process, this parameterization functions as a dynamical mapping: for any sampled point $\{A^\alpha_s, n^\alpha_s, r^\alpha\}$, {{}which are assigned priors from \autoref{table_priors_adapted}, the consistency relations (Eqs.~\labelcref{eq:consistency_1,eq:consistency_2,eq:consistency_3})} are used to determine the corresponding potential scale $V_0$, the deformation parameter $\alpha$, and the $e$-fold number $N_*$. {{}These parameters are subsequently read by \texttt{CLASS} as input to compute the primordial power spectra exactly. Once \texttt{CLASS} produces the spectra, the likelihood is evaluated by comparing the resulting $C_{\ell}$ with the data.} It is important to emphasize that while the parameterized variables $\{A^\alpha_s, n^\alpha_s, r^\alpha\}$ serve as efficient ``seeds'' for the sampling process, the final reported cosmological observables (e.g., $A_s, n_s, r$) are those {{}derived} numerically by \texttt{CLASS} {{}from the full dynamics}. This ensures that the results are not limited by the slow-roll approximation, but are instead derived from the exact dynamics of the potential. We further validate and discuss the parametrization against \texttt{CLASS} predictions in \autoref{app:A}. {{}The difference between the seed slow-roll values and the exact numerical observables can be appreciable, as we illustrate in \autoref{fig:consistency}.}

Thus, the cosmological parameter space for our analysis is spanned by the vector:

\begin{equation}
\mathbf{\Theta} = \{\omega_{\rm b}, \omega_{\rm cdm}, \tau_{\rm reio}, 100 \, \theta_{s}, \log(10^{10} A_s^\alpha), n_s^\alpha, r^\alpha\} \, . 
\label{eq:data_vector}
\end{equation}

Here, $\omega_{\rm b} \equiv \Omega_{\rm b}h^2$ and $\omega_{\rm cdm} \equiv \Omega_{\rm cdm}h^2$ represent the baryon and cold dark matter physical densities parameter respectively, while $\theta_s$ denotes the angular size of the sound horizon at recombination. {{}The parameter $\tau_{\rm reio}$ is the Thomson optical depth due to reionization.}

To robustly constrain the parameter space, our analysis incorporates a diverse suite of cosmological datasets: Baryon Acoustic Oscillation (BAO) measurements from the DESI Data Release 2 (DR2), CMB data from the Atacama Cosmology Telescope (ACT) Data Release 6, and the complete Planck 2018 legacy release. This multi-probe approach allows us to evaluate the slow-roll parameterization against the most precise observational bounds currently available.

\begin{itemize}
    \item[$\star$] \textbf{Baryon Acoustic Oscillations (DESI DR2):} 
    To map the cosmic expansion history, we incorporate the consensus Baryon Acoustic Oscillation (BAO) dataset from the DESI Data Release 2 \cite{2025PhRvD.112h3515A}. Extending across a redshift interval of $0.295 \leq z \leq 2.33$, this dataset utilizes galaxy and quasar tracers to provide geometric measurements of the transverse comoving distance, $D_M(z)/r_d$, and the Hubble distance, $D_H(z)/r_d$. Both quantities are normalized by $r_d$, the sound horizon at the drag epoch.
    
    \item[$\star$] \textbf{Primary CMB Anisotropies (Planck 2018):} 
    We adopt the Planck 2018 legacy release \cite{2020A&A...641A...5P} to constrain the primordial power spectrum via temperature and polarization anisotropies. Our analysis includes the auto and cross power spectra for temperature (T) and polarization (E). specifically, we use the $\tt{clik}$-based likelihoods for high multipoles ($30 \leq \ell \leq 2508$) in the TT, TE, and EE channels, complemented by the low-$\ell$ ($2 \leq \ell \leq 29$) TT and EE likelihoods. Furthermore, the priors over the nuisance parameters of the CMB likelihoods remain unaltered.

    \item[$\star$] \textbf{CMB Lensing (ACT DR6):} We probe the late-time growth of structure using the gravitational lensing convergence power spectrum, $C_\ell^{\kappa\kappa}$, from the Atacama Cosmology Telescope (ACT) Data Release 6 \cite{2024ApJ...962..113M,2024ApJ...962..112Q}. Extracted from high-resolution temperature and polarization maps (2017--2021), this dataset is integrated into our pipeline using the \texttt{act\_dr6\_lenslike} likelihood. 
\end{itemize}

The posterior distributions were obtained using the MCMC \cite{2002PhRvD..66j3511L} sampler within \texttt{Cobaya}. In order to guarantee robust parameter estimation, we evaluate the convergence of our sampling chains using the Gelman-Rubin diagnostic \cite{1992StaSc...7..457G}. Specifically, we impose a convergence threshold, normally requiring $R-1 < 0.01$ across all sampled parameters for every independent run. Finally, the marginalized constraints and contours were produced using the \texttt{GetDist} \cite{2025JCAP...08..025L}.

\section{Constraints from Planck, ACT and DESI}
\label{sec:constraints}

In this section we aim to put constraints on the Starobinsky and $\alpha$-Starobinsky inflationary scenarios with the aforementioned datasets. We first establish a baseline using the geometrically rigid Starobinsky model ($\alpha=1$) as a pipeline validation and to quantify the impact of adding ACT and DESI to the Planck constraints. We then lift this rigidity by allowing $\alpha$ to vary, enabling a broader exploration of inflationary dynamics directly informed by data. Throughout, we adopt the slow-roll parameterization of \autoref{subsec:consistency_relations} to map the observational seeds $\{A_s^\alpha, n_s^\alpha, r^\alpha\}$ into the underlying potential parameters, and we compute the final observables with full numerical integration in \texttt{CLASS}, avoiding reliance on slow-roll formulae at the level of predictions. 

\subsection{Starobinsky model, case \texorpdfstring{$\alpha=1$}{alpha=1}}
\label{subsec:Staro}

We treat this initial analysis as an essential sanity check. We begin by benchmarking our inference pipeline against the standard limit with the Starobinsky model ($\alpha=1$). We can verify that our numerical pipeline behaves as expected before introducing the added flexibility of the $\alpha$ deformation. For this specific case, we present results for two configurations: a Planck-only and a joint analysis including the full suite of Planck, ACT, and DESI measurements.\\ 
By fixing $\alpha = 1$, the slow-roll parameterization reduces the inflationary degrees of freedom such that the mapping depends solely on the potential scale $V_0$ and the number of $e$-folds $N_*$. Consequently, the sampling is effectively driven by the observational set $\{A_s^\alpha, n_s^\alpha\}$ via the Starobinsky consistency relation. Following our general framework, these ``seed'' values are mapped to the potential parameters for full numerical integration in \texttt{CLASS}. Thus, the resulting inflationary observables, specifically $A_{s}$, $n_{s}$, and $r$, are numerically computed without relying on standard slow-roll approximations.  

\begin{figure}[htbp]
    \centering
    \includegraphics[width=\textwidth]{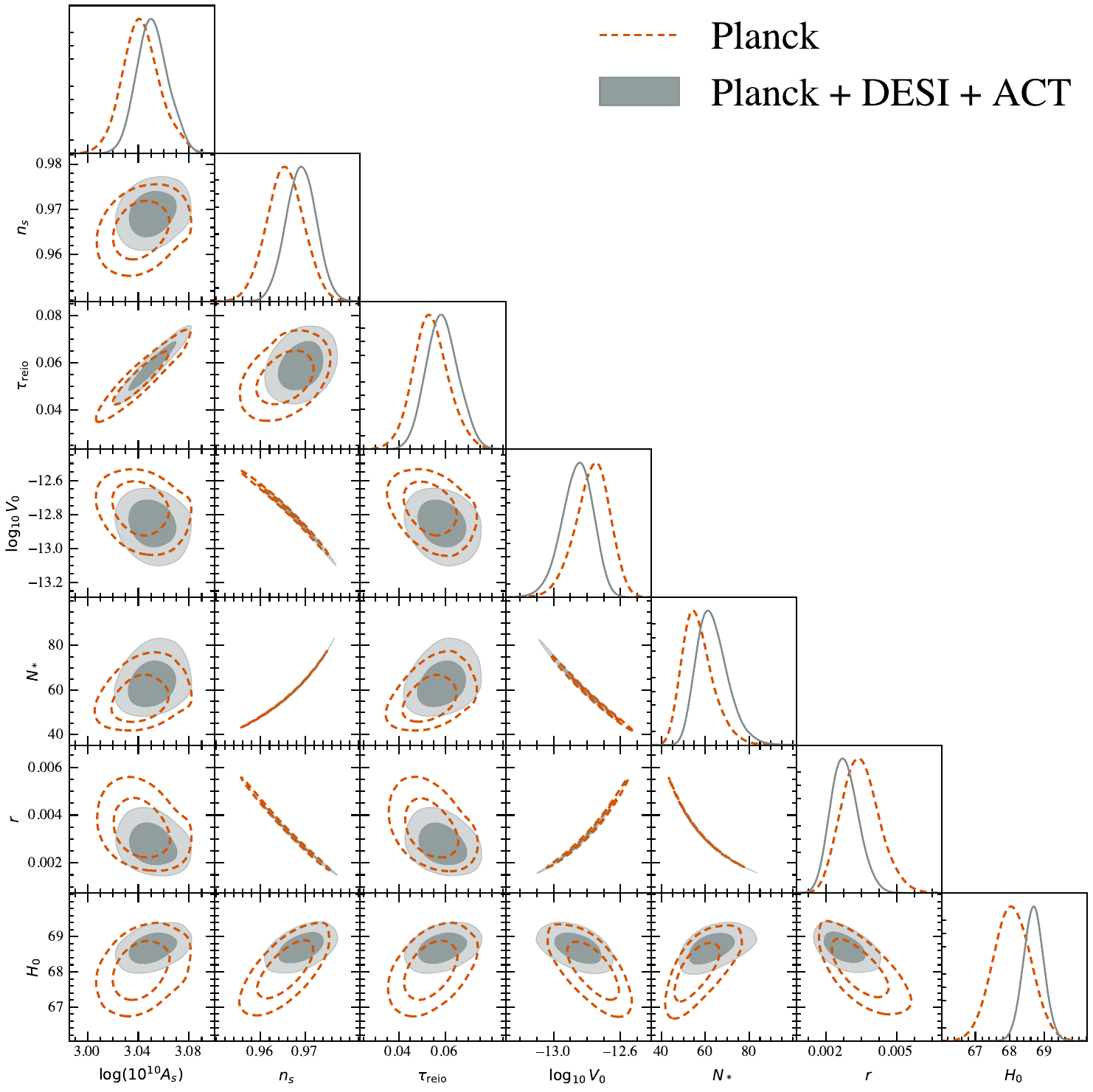}
    \caption{Posterior constraints for the Starobinsky model ($\alpha=1$). One- and two-dimensional marginalized distributions compare Planck-only (dashed orange) with the joint Planck+ACT+DESI analysis (solid gray). The geometrical rigidity of the model enforces a tight correlation between $n_s$ and $r$. Incorporating DESI and ACT shifts the preferred tilt toward higher mean value ($n_s = 0.969$) and, within the $\alpha=1$ track, this is accommodated by a larger mean value of $e$-folds ($N_* = 63.4$), pushing against the known theoretical canonical $50 \leq N_* \leq 60$ range.}
    \label{fig:staro_constraints}
\end{figure}

The results for the Starobinsky model ($\alpha=1$) are presented in \autoref{fig:staro_constraints} and summarized in \autoref{table_starobinsky_baseline}. A primary finding is the notable shift in the preferred inflationary dynamics when moving from the Planck-only to the combined Planck+DESI+ACT analysis. Specifically, the full dataset favors a higher scalar spectral index, $n_s \approx 0.97$, compared to the Planck result of $0.9654 \pm 0.0041$. {{}This shift is partly driven by the well-known correlation between $n_s$ and $H_0$: DESI favors a higher Hubble constant than Planck alone, and this preference is translated into higher values of $n_s$ through their parameter degeneracy.} Because the $\alpha=1$ model is geometrically rigid, this increase in $n_s$ is directly coupled to the number of $e$-folds $N_*$. Consequently, the posterior for $N_*$ shifts from $56.7^{+5.2}_{-8.3}$ to a much higher value of $63.4^{+5.5}_{-8.2}$ to maintain consistency with the data. \\
This shift yields an unsuitable realization within the Starobinsky scheme since it is increasingly pushed beyond the standard theoretically motivated range ($50 < N_* < 60$) for the duration of inflation (as previously reported in \cite{2025arXiv251106640B}). Furthermore, we observe a narrow, linear degeneracy between $n_s$ and $r$, confirming that for a fixed $\alpha$, the model can only slide along a specific theoretical track. These results provide the motivation for our subsequent analysis of the $\alpha$-deformed case, where the additional degree of freedom allows the model to accommodate the diverse datasets without requiring such extreme values for $N_*$. 

\begin{table}[htbp]
\centering
\setlength{\tabcolsep}{10pt} 
\small 
\renewcommand{\arraystretch}{1.5} 
\begin{tabular}{l c c }
\toprule
\textbf{Parameter} & \textbf{Planck} & \textbf{Planck+ACT+DESI} \\
\midrule

$\Omega_b h^2$  
& $0.02239\pm 0.00014$  
& $0.02251\pm 0.00013$   
\\

$\Omega_c h^2$  
& $0.1196\pm 0.0012$  
& $0.11827\pm 0.00064$   
\\

$\tau_\mathrm{reio}$  
& $0.0534\pm 0.0074$  
& $0.0584\pm 0.0066$   
\\

$\log_{10}(V_0)$  
& $-12.77^{+0.11}_{-0.089}$  
& $-12.86^{+0.10}_{-0.083}$   
\\

$N_{*}$  
& $56.7^{+5.2}_{-8.3}$  
& $63.4^{+5.5}_{-8.2}$   
\\

$H_{0} \, [\text{km}\,\text{s}^{-1}\,\text{Mpc}^{-1}]$  
& $68.06\pm 0.54$  
& $68.68\pm 0.30$   
\\

$\log(10^{10} A_\mathrm{s})$
& $3.042\pm 0.014$  
& $3.051\pm 0.012$   
\\

$n_{s}$  
& $0.9654\pm 0.0041$  
& $0.9690\pm 0.0033$   
\\

$r$  
& $0.0035^{+0.0016}_{-0.0015}$
& $0.0028^{+0.0012}_{-0.0011}$   
\\

\bottomrule
\end{tabular}
\caption{Constraints on cosmological and derived inflationary parameters for the Starobinsky model ($\alpha=1$), comparing Planck-only with the combined Planck+ACT+DESI analysis. The joint dataset favors a slightly higher tilt $n_s$ and, along the rigid $\alpha=1$ track, a larger $N_*$, illustrating an inconsistency with the canonical $N_*\sim50$–$60$ theoretical window. Uncertainties are quoted at $1\sigma$; $H_0$ is in $\mathrm{km\,s^{-1}\,Mpc^{-1}}$.}
\label{table_starobinsky_baseline}
\end{table}

\subsection{\texorpdfstring{$\alpha$}{alpha}-Starobinsky model}
\label{subsec:alpha_staro}

For our second phase of analysis, we added flexibility of the deformation parameter to test the broader $\alpha$-Starobinsky scheme. Here, the core parameters governing the inflationary potential are sampled concurrently with the standard cosmological degrees of freedom, using various configurations of the Planck, ACT, and DESI measurements.

\begin{figure}[htbp]
\centering
  \includegraphics[width=\linewidth]{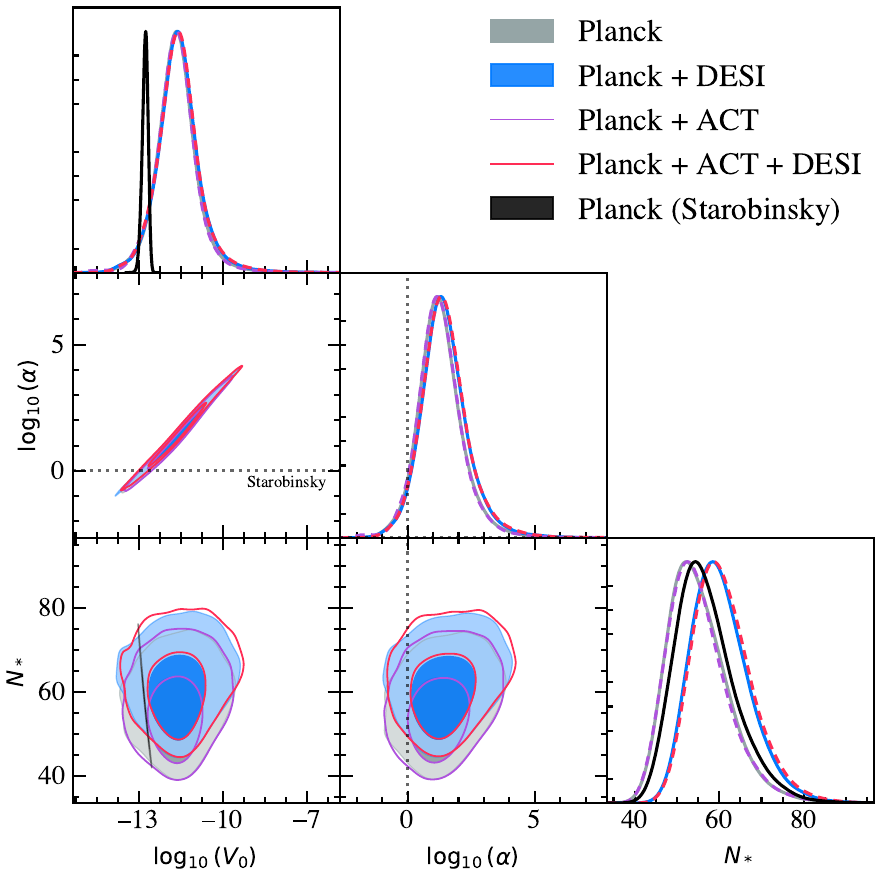} \caption{Marginalized 1D and 2D posteriors for $\{\log_{10} V_0, \log_{10}\alpha, N_*\}$ in the $\alpha$-Starobinsky model. Contours (68\% and 95\% credible regions) are shown for Planck (gray), Planck+DESI (blue), Planck+ACT (purple), and the full combination (red). Dotted lines indicate the Starobinsky limit $\log_{10}\alpha=0$. While $\alpha=1$ remains compatible at $2\sigma$, the joint dataset disfavors it at $1\sigma$ and prefers $\log_{10}\alpha>0$, indicating a mild but data-driven tilt toward a broader plateau. The near overlap of Planck and Planck+ACT contours shows that ACT adds little constraining power for these specific inflationary parameters beyond Planck.}
  \label{fig:constraints_key_params_alpha}
\end{figure}

As observed in the marginalized posteriors of \autoref{fig:constraints_key_params_alpha}, the Starobinsky model ($\alpha=1$) is pushed outside the $1\sigma$ allowed region by each analysis. The quantitative results of the rest of parameters are summarized in \autoref{table_results_combined}, where the bounds for each cosmological parameter is reported at the $1\sigma$ confidence level, while the constraints on $r$ are given as $95\%$ CL upper limits. For completeness, the full set of 1D and 2D marginalized posterior distributions for the entire $\alpha$-Starobinsky parameter space is provided in \autoref{fig:alpha_full_constraints} in the \autoref{app:B}.

\begin{figure}[htbp]
\centering
  \includegraphics[width=\linewidth]{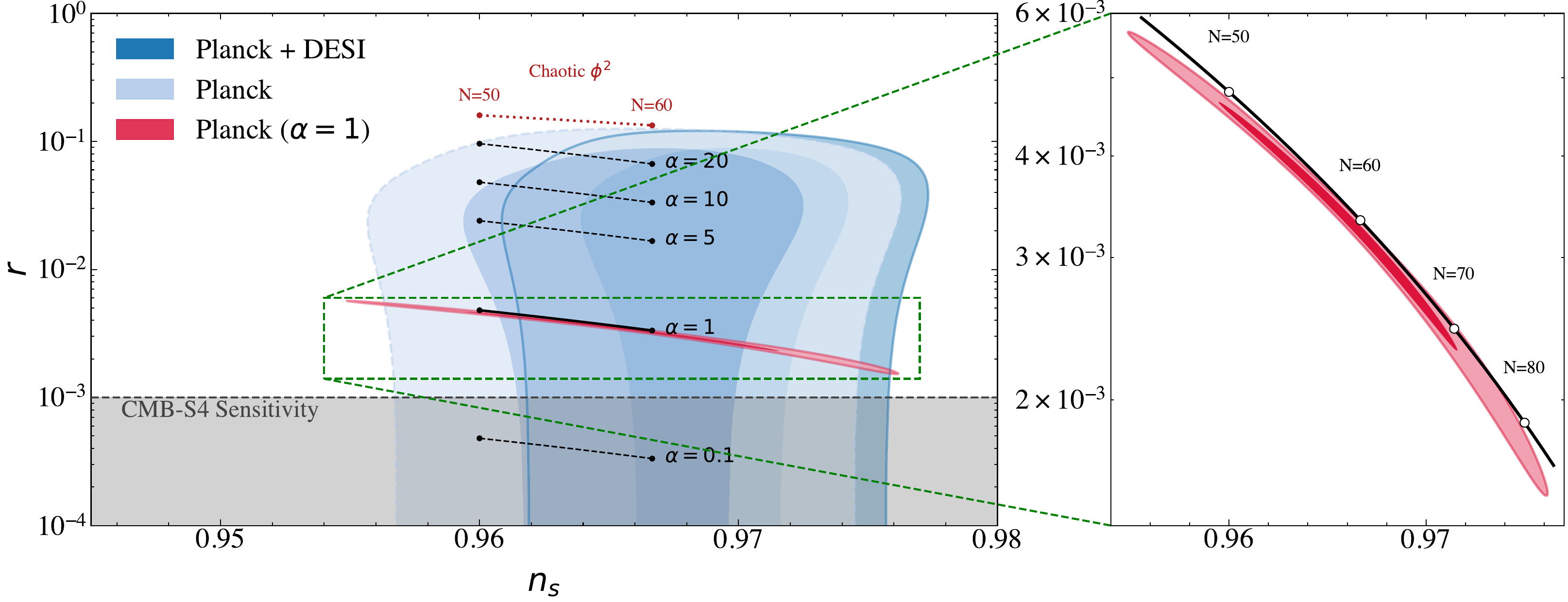}
  \caption{Joint 68\% and 95\% CL constraints in the $(n_s, r)$ plane at $k=0.05\,\mathrm{Mpc}^{-1}$ for Planck (light blue) and Planck+DESI (dark blue). The red contour shows the Planck constraint restricted to the $\alpha=1$ model. Theoretical $\alpha$-Starobinsky tracks (black) are overlaid for selected $\alpha \in \{0.1, 1, 5, 10, 20\}$; filled circles mark $N=\{50,60\}$, and the zoom highlights the $\alpha=1$ locus with open circles for $N=\{50,60,70,80\}$. {{}DESI drives the preferred $n_s$ to higher values. In the $\alpha=1$ case, this forces $N$ to larger values along the fixed track; by contrast, the full $\alpha$-Starobinsky model accommodates it by adopting values $\alpha > 1$, which raises $r$ at fixed $N$ instead.}
  The gray band ($r<10^{-3}$) indicates the target sensitivity of future CMB-S4 \cite{2019arXiv190704473A,2022ApJ...926...54A}. {{}Noticing, the $\alpha=1$ contour is tightly constrained by the reduced parameter space and largely falls outside the CMB-S4 sensitivity band, while the broader $\alpha$-Starobinsky contours remain consistent with it.} Large-$N$ slow-roll estimates are used for the theory tracks: $n_s \simeq 1-2/N$ and $r \simeq 12\alpha/N^2$ (and $r \simeq 8/N$ for quadratic chaotic inflation).}
  \label{fig:constraints_r_ns}
\end{figure}

Furthermore, as illustrated in the $n_s - r$ plane (\autoref{fig:constraints_r_ns}), the inclusion of DESI data visibly shifts the preferred scalar spectral index to higher values compared to the Planck-only analysis.
{{}As in the Starobinsky limit discussed in \autoref{subsec:Staro}, this shift is partly driven by the $n_s$--$H_0$ correlation: DESI's preference for a larger Hubble constant translates into higher values of $n_s$. The same trend persists here; however, the additional freedom of the deformation parameter $\alpha$ allows the model to accommodate this shift without resorting to an unsuitable number of $e$-folds, in contrast to the rigid $\alpha=1$ case.} To maintain consistency with this latest BAO data, the model strongly prefers a positive deformation ($\log_{10} \alpha > 0$) to raise the tensor-to-scalar ratio, or an undesirably large number of $e$-folds ($N_* \gtrsim 70$) to shift $n_s$ higher along the standard $\alpha=1$ track. 

\begin{figure}[htbp]
\centering
  \includegraphics[width=\linewidth]{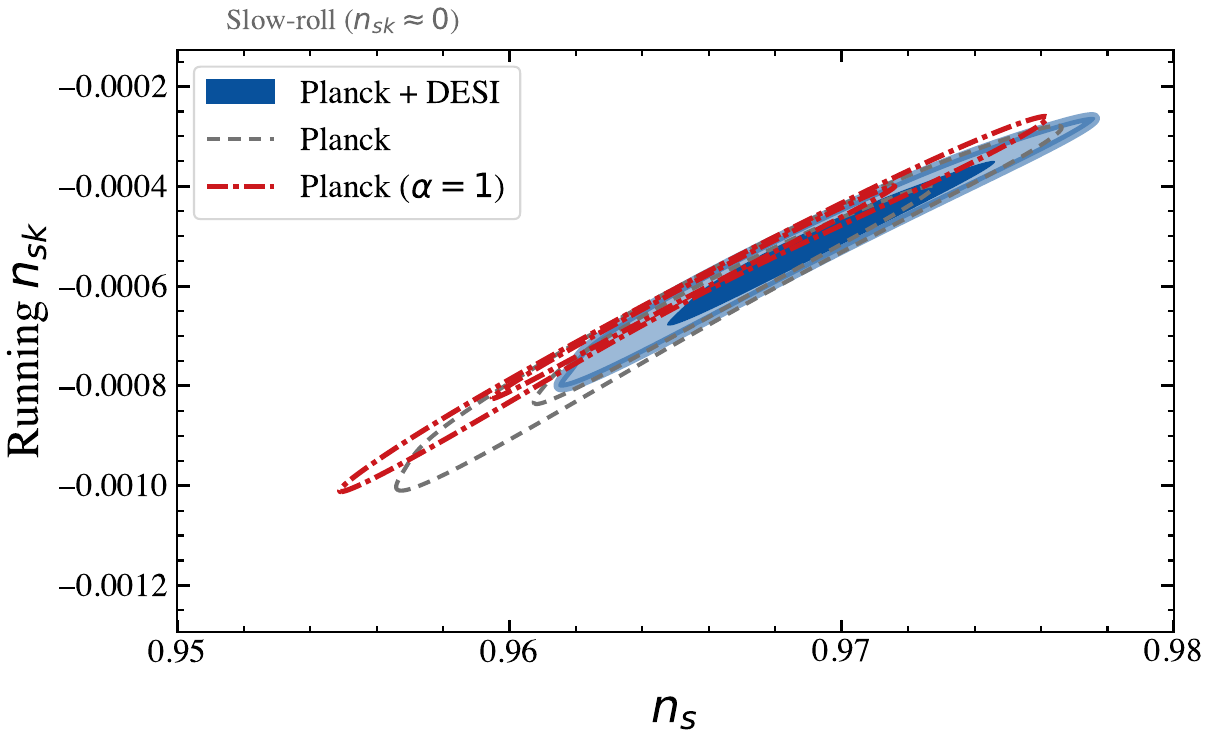}
  \caption{Marginalized $68\%$ and $95\%$ confidence regions for the scalar running $n_{sk} \equiv d n_s / d \log k$ as a function of the scalar spectral index $n_s$. The pure Planck posterior (dashed gray) is compared against the joint Planck + DESI constraints (filled blue). The inclusion of BAO information preserves the overall correlation between the two parameters but distinctly shifts the preferred region toward higher $n_s$, as expected. Both dataset combinations exhibit a preference for a small, negative running ($n_{sk} < 0$), though the standard slow-roll benchmark ($n_{sk} \approx 0$, denoted at top in gray font) remains highly consistent with the data. Moreover, the pure Planck case of the canonical Starobinsky case ($\alpha=1$, red dash-dot line) is overlaid, illustrating again how the DESI displacement places the strict $\alpha=1$ limit under increasing observational pressure.}  \label{fig:nskns}
\end{figure}

To further examine the scale dependence of the primordial perturbations, \autoref{fig:nskns} presents the derived contours for the $n_{s} - n_{sk}$ plane. These predictions are obtained using the analytical consistency relation for the scalar running presented in Eq.~\eqref{eq:nsk1}. Consistent with our earlier findings for the scalar tilt, the inclusion of DESI data systematically shifts the two-dimensional contour toward higher values of $n_{s}$ relative to the pure Planck analysis, due to the underlying $n_{s}$ dependence. Notably, both dataset combinations strictly bound the scalar running to small, negative values ($n_{sk} < 0$), which aligns perfectly with second-order slow-roll expectations. Crucially, examining the Starobinsky limit ($\alpha=1$) within this parameter space reinforces our previous conclusions: the rightward displacement driven by DESI pushes the canonical model to the periphery of the $68\%$ confidence region. This subtle but persistent geometric strain on the standard $R^2$ scenario further justifies the necessity of the broader $\alpha$-Starobinsky framework.

\begin{table}[htbp]
\centering
\setlength{\tabcolsep}{4pt} 
\scriptsize 
\renewcommand{\arraystretch}{1.5} 
\begin{tabular}{l c c c c }
\toprule
\textbf{Parameter} & \textbf{Planck} & \textbf{Planck+ACT} & \textbf{Planck+DESI} & \textbf{Planck+ACT+DESI} \\
\midrule

$\Omega_b h^2$ 
& $0.02240\pm 0.00015$ 
& $0.02239\pm 0.00015$ 
& $0.02249\pm 0.00013$ 
& $0.02250\pm 0.00013$ 
\\

$\Omega_c h^2$ 
& $0.1195\pm 0.0012$ 
& $0.1196\pm 0.0012$ 
& $0.11824\pm 0.00065$ 
& $0.11827\pm 0.00064$ 
\\

$\tau_\mathrm{reio}$ 
& $0.0535\pm 0.0073$ 
& $0.0539\pm 0.0074$ 
& $0.0570\pm 0.0071$ 
& $0.0580\pm 0.0067$ 
\\

$\log_{10}(V_0)$ 
& $-11.61\pm 0.76$ 
& $-11.66\pm 0.79$ 
& $-11.55^{+0.62}_{-0.73}$ 
& $-11.56\pm 0.82$ 
\\

$\log_{10}(\alpha)$ 
& $1.32^{+0.69}_{-0.91}$ 
& $1.27^{+0.72}_{-0.87}$ 
& $1.48^{+0.72}_{-0.96}$ 
& $1.47^{+0.73}_{-0.95}$ 
\\

$N_{*}$ 
& $55.2^{+5.1}_{-8.3}$ 
& $55.1^{+5.0}_{-8.3}$ 
& $60.5^{+5.2}_{-8.0}$ 
& $61.0^{+5.3}_{-8.1}$ 
\\

$H_{0} \, [\text{km}\,\text{s}^{-1}\,\text{Mpc}^{-1}]$
& $68.08\pm 0.55$ 
& $68.03\pm 0.53$ 
& $68.68\pm 0.29$ 
& $68.67\pm 0.30$ 
\\

$\log(10^{10} A_\mathrm{s})$
& $3.041\pm 0.014$ 
& $3.043\pm 0.013$ 
& $3.046\pm 0.014$ 
& $3.050\pm 0.012$ 
\\

$n_{s}$ 
& $0.9662\pm 0.0042$ 
& $0.9661\pm 0.0042$ 
& $0.9692\pm 0.0033$ 
& $0.9695\pm 0.0033$ 
\\

$r$
& $< 0.1$ 
& $< 0.098$  
& $< 0.097$  
& $< 0.096$  
\\

\bottomrule
\end{tabular}
\caption{Marginalized $1\sigma$ constraints on standard cosmological parameters and the primary $\alpha$-Starobinsky set $\{\log_{10}V_0,\log_{10}\alpha,N_*\}$ for Planck-only, Planck+ACT, Planck+DESI, and the full combination. Upper limits on $r$ are quoted at 95\% CL. The DESI BAO information primarily drives the upward shift in $n_s$ and the preference for $\log_{10}\alpha>0$.}
\label{table_results_combined}
\end{table}

\section{Conclusion and final remarks}
\label{sec:conclusions}

In this work, we have revisited the viability of the $\alpha$-Starobinsky inflationary model. By implementing the model within the \texttt{CLASS} Boltzmann solver, we explored its specific imprints on the CMB anisotropies in \autoref{fig:TT_impact}, demonstrating that it can be robustly constrained by current observations. To this end, we probed the model using a high-precision joint dataset comprising Planck 2018, ACT DR6 CMB lensing, and DESI DR2 BAO. A central approach of our analysis was the implementation of a novel priors parameterization strategy during parameter inference. By placing priors directly on the empirically accessible primordial observables $(A_s, n_s, r)$ and mapping them via analytical consistency relations, we reconstructed the physical parameters of the $\alpha$-Starobinsky potential $(V_0, \alpha, N_*)$. The exact background and perturbation dynamics were subsequently evolved using a modified \texttt{CLASS} solver. This ensured that our posterior constraints remain numerically rigorous and anchored to observational bounds rather than abstract energy scales. \\
Our analysis of the canonical Starobinsky model ($\alpha=1$) reveals a pronounced friction driven by the latest large-scale structure measurements. The inclusion of BAO footprint through DESI data systematically shifts the preferred scalar spectral index toward higher values ($n_s \approx 0.97$), as shown in \autoref{fig:staro_constraints}. Because the pure Starobinsky potential is geometrically rigid, accommodating this spectral tilt strictly dictates the duration of inflation, forcing the required a mean value of $e$-folds to $N_* = 63.4$. This firmly pushes the canonical model beyond the theoretically well-motivated window of $50 < N_* < 60$, signalling that the standard scenario struggles to naturally align with the contemporary cosmic expansion history. \\
This unsuitable realization provides a compelling justification for extending the analysis into the generalized $\alpha$-Starobinsky scheme. By allowing the deformation parameter $\alpha$ to vary, we effectively decouple the width of the inflationary plateau from the strict $\alpha=1$ limit. Our joint analysis demonstrates that this additional degree of freedom successfully absorbs the observational strain. As illustrated in \autoref{fig:constraints_key_params_alpha}, not only does each dataset exhibit a clear statistical preference for a broader plateau ($\log_{10} \alpha > 0$), but this shift is substantial enough that the canonical Starobinsky model ($\alpha=1$) is pushed entirely outside the $1\sigma$ allowed region across all analyses. \\ 
Moreover, as illustrated in the $n_{s} - r$ plane (see \autoref{fig:constraints_r_ns}), this extended parameter space adjusts the higher scalar tilt, $n_{s}$, favored by DESI. Crucially, allowing $\alpha$ variation breaks the strict parameter degeneracy of the pure Starobinsky model, relaxing the upward pull on the number of $e$-folds and shifting its central value lower ($N \simeq 61$) to align more comfortably with theoretical expectations. Furthermore, although incorporating high-resolution CMB lensing data from ACT DR6 serves as an excellent consistency check, its constraining power on the primordial parameter space is marginal; the resulting posteriors are predominantly driven by the synergy between Planck and DESI, with the latter primarily responsible for the shift in $n_{s}$. Finally, the derivation of the scalar running ($n_{sk}$) further solidifies our findings. 

The analysis developed here serves for several potential avenues of future work. First, the current parameter space can be further constrained by incorporating explicit reheating information \cite{2014PhRvL.113d1302D,2015PhRvD..91d3521M,2024EPJP..139..302G} into the Bayesian analysis. Linking the post-inflationary equation of state $\omega_{\rm rh}$ and the reheating temperature $T_{\rm rh}$ might help to break residual degeneracies and place bounds on the Reheating parameters. Second, to overcome the computational bottleneck of the Boltzmann solver by exactly solving the inflationary dynamics at every MCMC step, we plan to implement neural emulators (e.g. using \texttt{CosmoPower} \cite{2022MNRAS.511.1771S}) to drastically accelerate-up the analysis without sacrificing numerical precision. With enhanced computational efficiency, it will become feasible to perform robust Bayesian model comparisons, statistically weighing the $\alpha$-Starobinsky paradigm against other compelling inflationary models (see e.g. \cite{2025JCAP...03..043S}), such as the broader family of generalized $\alpha$-attractors. 
Finally, to extend our analysis, we will leverage cutting-edge computational frameworks, such as the fully differentiable likelihood pipeline \texttt{candl} \cite{2024A&A...686A..10B} (which natively has likelihoods for Planck, ACT, and the South Pole Telescope). In addition, integrating gradient-based methods will substantially enhance our sampling efficiency, ensuring precise constraints on inflationary models as we prepare for the next generation of precision cosmology.

\section*{Data Availability}
\phantomsection
\label{data_availability}

The data underlying this article will be shared on reasonable request to the corresponding author.

\section*{Acknowledgments}
\phantomsection
\label{acknowledgments}

KC and FXLC acknowledge the postdoctoral fellowship from the Secretaría de Ciencias, Humanidades, Tecnología e Innovación (SECIHTI). GG and JCH would like to thank DGAPA-PAPIIT-UNAM grant No. IN110325  \textit{``Estudios en cosmolog\'ia inflacionaria, agujeros negros primordiales y energ\'ia oscura"} and JCH acknowledges grant No. IN114626 \textit{``Estudios de ANPs y Contribuciones In-Kind a la colaboración LSST-México"}. The authors gratefully acknowledge the computing facilities provided by the Atocatl cluster at LAMOD-UNAM. LAMOD (\url{http://www.lamod.unam.mx/}) is a collaborative effort between the IA, ICN, and IQ institutes at UNAM. {{}We also thank the anonymous referee for their valuable comments and suggestions, which helped improve this manuscript.}

\appendix

\section{Validation of consistency relations}
\label{app:A}

\begin{figure}[htbp]
\centering
  \includegraphics[width=\linewidth]{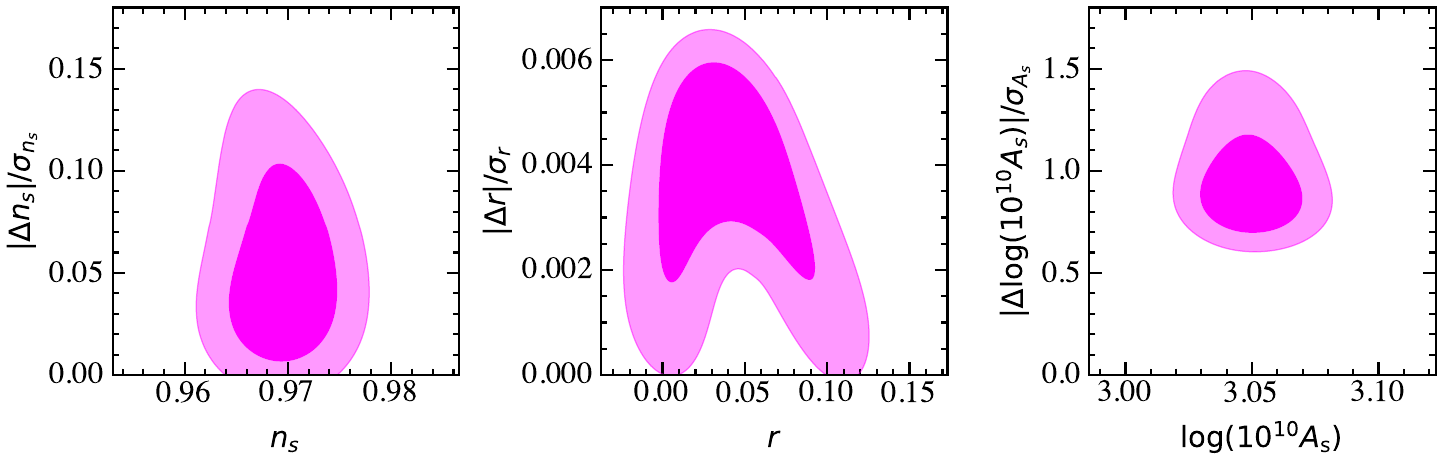}
  \caption{Validation of the slow-roll parameterization against full numerical predictions from \texttt{CLASS} for the joint Planck+ACT+DESI analysis. The vertical axes show the parameter differences normalized by the marginalized uncertainties, $\lvert \Delta X \rvert/\sigma_X= \vert  X^{\texttt{CLASS}} - X^{\alpha} \vert  / \sigma_X$ with $X\in\{n_s,r,\log(10^{10}A_s)\}$. We find excellent agreement for $n_s$ and $r$ (sub-$0.1\sigma$ shifts), while $\log(10^{10}A_s)$ exhibits a $\sim 1\sigma$ offset, consistent with missing next-to-leading order slow-roll corrections. Accordingly, we treat the parameterization as an efficient sampling prior and report final observables from the full numerical pipeline.}
  \label{fig:consistency}
\end{figure}

We have performed a consistency check between our analytical approximation (denoted as $\alpha$) and the full numerical Boltzmann solver \texttt{CLASS}. To quantify the agreement, we computed the posterior distribution of the parameter shifts normalized by their marginal standard deviations, defined as $\vert \Delta \theta \vert / \sigma_\theta = \vert \theta^{\texttt{CLASS}} - \theta^{\alpha}  \vert / \sigma_\theta$ which approximates how many $\sigma$ apart the predictions are.  \\
The results of this validation are shown in \autoref{fig:consistency}. For the spectral index $n_s$ and the tensor-to-scalar ratio $r$, we find excellent agreement between the two approaches. The systematic shifts are $\Delta n_s \approx 0.05\sigma$ and $\Delta r \approx 0.004\sigma$, which are negligible compared to the statistical precision of current data (Planck + ACT + DESI). This confirms that our approximation is sufficiently accurate for inferring inflationary tilt and tensor modes without introducing too much parameter bias.

On the other hand, for the scalar amplitude $\log(10^{10}A_s)$, we observe a systematic shift of $\Delta \log 10^{10} A_s \approx 0.94\sigma$. While the inferred contours overlap, this offset indicates that the leading-order slow-roll approximation underestimates the amplitude required to match the numerical solution at the pivot scale $k_* = 0.05 \, \text{Mpc}^{-1}$. This discrepancy is consistent with missing next-to-leading order (NLO) corrections in the slow-roll expansion (e.g., Stewart-Lyth corrections \cite{1993PhLB..302..171S}, which are known to be of order $\mathcal{O}(\epsilon) \sim$ percent-level \cite{1992PhR...215..203M,2009pdp..book.....L}). For analyses requiring high-precision constraints on $A_s$ or $S_8$, this approximation should be calibrated or extended to NLO; for constraints on $n_s$ and $r$, the current precision is robust. 

As emphasized throughout this work, our inference analysis imposes uniform priors directly on the primordial power spectrum parameters, rather than on the physical variables of the inflationary potential. This translation, governed by the analytical consistency relations, is summarized in \autoref{tab:priors_alpha_params}. Notably, this mapping reveals that a uniform prior in the primordial basis $\{ A_s^\alpha, n_s^\alpha, r^\alpha \}$ does not restrict the model parameters $\{V_0, \alpha, N_*\}$ to a finite domain; specifically, while the lower bounds are well-defined, the upper limits remain physically unbounded, allowing for a semi-infinite parameter space. Because the $(4\delta_{n_s} - r)$ term in the denominator of the exact relations can cross zero, the theoretical bounds for $V_0$, $N_*$, and $\alpha$ formally diverge.

\begin{table*}[htbp]
\centering
\renewcommand{\arraystretch}{1.4}
\begin{tabular}{ccc}
\hline 
\textbf{Parameter} & \textbf{Bounds} & \textbf{Consistency relation} \\ 
\hline\\[-2mm]
$\log_{10}(V_0/M_{\rm Pl}^4)$ & $[-20, \infty)$ & Eq.~\eqref{eq:consistency_2} \\[2mm]
$N_{*}$ & $[0, \infty)$ & Eq.~\eqref{eq:consistency_3} \\[2mm]
$\alpha$ & $[0, \infty)$ & Eq.~\eqref{eq:consistency_1} \\[2mm]
\hline 
\end{tabular}
\caption{Translation of the uniform  priors on the primordial parameters ($A^\alpha_s$, $n^\alpha_s$, and $r^\alpha$) into the corresponding parameters of the $\alpha$-Starobinsky model via the consistency relations. The theoretical bounds formally diverge because the term $(4\delta_{n_s} - r)$ in the denominator of the consistency relations smoothly crosses zero within the allowed prior volume.}
\label{tab:priors_alpha_params} 
\end{table*}

\section{Full posteriors of \texorpdfstring{$\alpha$}{alpha}-Starobinsky}
\label{app:B}

\begin{figure}[htbp]
\centering
\includegraphics[width=\linewidth]{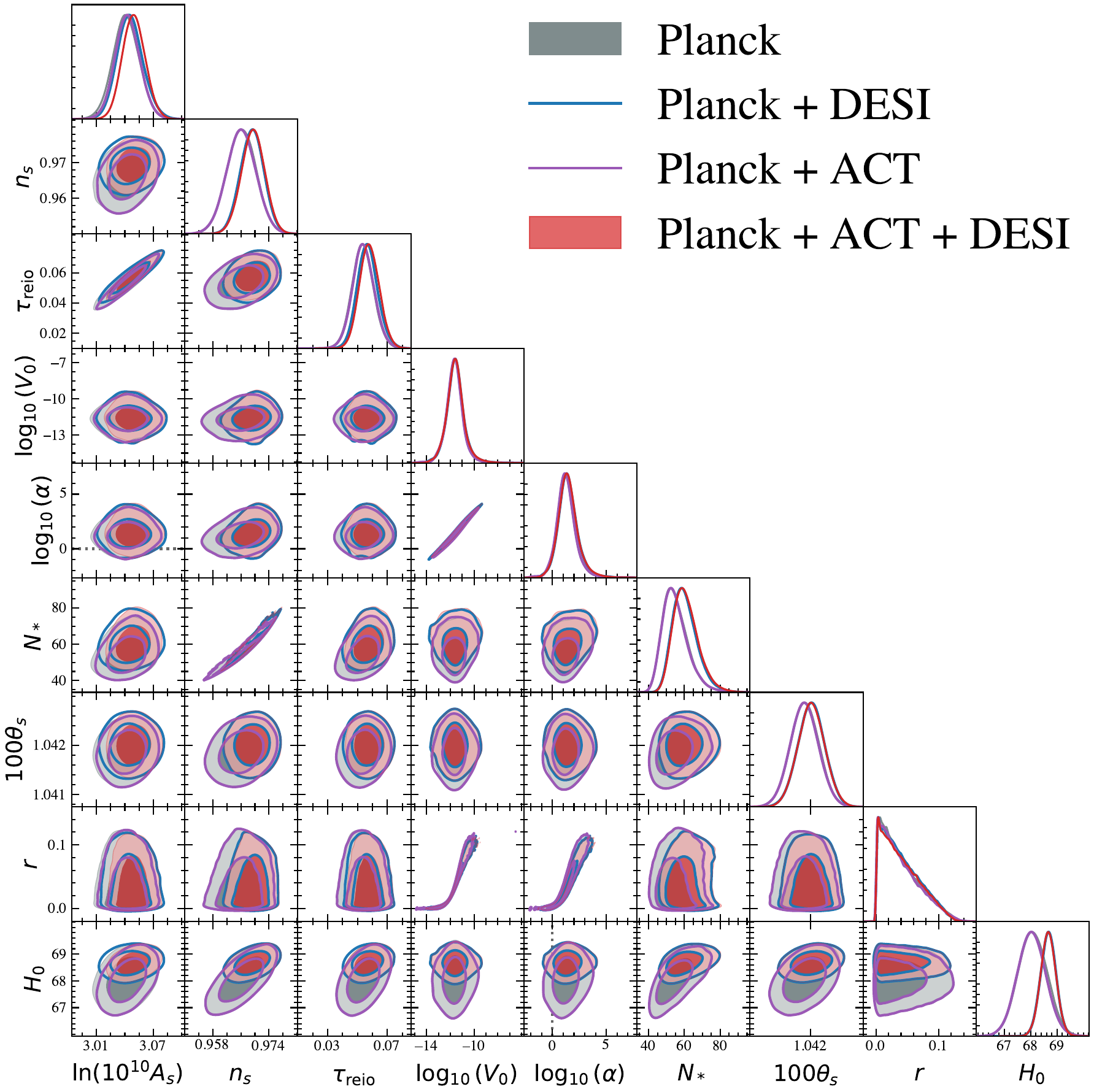}
\caption{Full 1D and 2D marginalized posteriors for the $\alpha$-Starobinsky analysis across all cosmological and inflationary parameters. Filled contours indicate Planck (gray) and the full Planck+ACT+DESI combination (red), with Planck+DESI (blue) and Planck+ACT (purple) overlaid. The vertical line at $\log_{10}\alpha=0$ marks the Starobinsky limit. This view exposes residual degeneracies and confirms that ACT lensing adds limited constraining power beyond Planck for $(\log_{10}V_0,\log_{10}\alpha,N_*)$, whereas DESI drives the shift toward higher $n_s$, thus impacting the $e$-fold. Note that the primordial observables $A_{s}$, $n_{s}$, and $r$ are computed directly via \texttt{CLASS}.}  
\label{fig:alpha_full_constraints}
\end{figure}

For completeness, this appendix provides the full triangle plot encompassing all sampled and derived cosmological parameters of the $\alpha$-Starobinsky model evaluated in this work. As illustrated in \autoref{fig:alpha_full_constraints}, this comprehensive view explicitly maps the residual parameter degeneracies and visually reinforces the primary conclusions discussed in the main text. In particular, the filled contours display the individual and combined constraints from Planck (gray) and the full Planck+ACT+DESI data set (red), with Planck+DESI (blue) and Planck+ACT (purple) overlaid for comparison. The vertical reference at $\log_{10}\alpha=0$ highlights the Starobinsky limit, providing a convenient guide to assess departures favored by the data. The posteriors reveal that ACT lensing contributes only modest additional constraining power beyond Planck in the inflationary subspace $(\log_{10}V_0,\log_{10}\alpha,N_*)$, while DESI mainly drives a shift toward higher $n_s$, which in turn propagates into the inferred number of $e$-folds $N_*$. This behavior clarifies the structure of the remaining degeneracies and the relative role of each data set in tightening parameter constraints across the cosmological inflationary parameter block.

\clearpage 

\bibliographystyle{JHEP}
\bibliography{bib}

\end{document}